\documentclass[12pt,twocolumn]{article}
\usepackage{amsmath,latexsym}
\usepackage{graphicx}
\begin{document}
\title{\textbf{\textsf{ Quintom Wormholes }}}
\author{
\large Peter K. F. Kuhfittig \footnote{kuhfitti@msoe.edu}, \large
Farook Rahaman\footnote{farook\_rahaman@yahoo.com}  and
 Ashis Ghosh \footnote{ashis\_laru@yahoo.co.in} \\ \\
$^\dag$\small Department of Mathematics, Milwaukee School of
Engineering,
Milwaukee, Wisconsin 53202-3109, USA\\
$^\ddag$\small Department of Mathematics, Jadavpur University,
Kolkata - 700032,
 India\\
\small \\
 }\maketitle
 \begin{abstract}
The combination of quintessence and phantom energy in a joint
model is referred to as quintom dark energy.  This paper
discusses traversable wormholes supported by such quintom
matter.  Two particular solutions are explored, a constant
redshift function and a specific shape function.  Both
isotropic and anisotropic pressures are considered.

\end{abstract}

  \footnotetext{

}
    \mbox{} \hspace{.2in}

\section{ {\textbf{ Introduction}} }

A stationary spherically symmetric wormhole may be defined as a
handle or tunnel in a multiply-connected spacetime joining widely
separated regions of the same spacetime or of different
spacetimes \cite{MT88}.  Interest in wormholes was renewed by
the realization that much of our Universe is pervaded by a dynamic
dark energy that causes the Universe to accelerate \cite{aR98,
sP99}: $\overset{..}{a}$ is positive in the Friedmann equation
$\overset{..}{a}/a=-\frac{4\pi G}{3}(\rho+3p)$.  In the equation
of state (EoS) $p=\omega\rho$, a value of $\omega <-1/3$ is
required for an accelerated expansion.  The range of values
$-1<\omega<-1/3$ is usually referred to as \emph{quintessence}
and the range $\omega<-1$ as \emph{phantom energy}.  In the latter
case we get $\rho+p<0$, in violation of the weak energy condition,
considered to be a primary prerequisite for the existence of
wormholes \cite{oZ05, fL05, pK06}.  The special case
$\omega=-1$ corresponds to Einstein's cosmological constant.
This value is sometimes called the \emph{cosmological constant
barrier} or the \emph{phantom divide}.

The quintessence and phantom energy models taken together as
a \emph{joint model} and dubbed \emph{quintom} for short
\cite{FWZ05} suggests a single EoS, $p=\omega\rho$, to cover
all cases.  It is shown in Ref. \cite{XCQZZ}, however,
that $\omega$ cannot cross the phantom divide, that is,
in the traditional scalar field model, the EoS cannot cross
the cosmological constant barrier.  (For further discussion,
see Refs. \cite{ZXLFZ, CD05, aV05, wH05, KS06}.)  The
simplest quintom model involves two scalar fields, $\phi$
and $\psi$, one quintessence-like and one phantom-like
\cite{CQBPZ, GPZZ}:
\begin{multline*}
   \mathcal{L}=\frac{1}{2}\partial_{\mu}\phi
       \partial^{\mu}\phi
     -\frac{1}{2}\partial_{\mu}\psi
       \partial^{\mu}\psi\\-V(\phi)-W(\psi).
\end{multline*}
So instead of $\omega$ in the EoS, we will use
$\omega_q$ and $\omega_{ph}$, respectively.

One reason for studying quintom dark energy is the bouncing
universe \cite{CQBPZ, CQPLZ, ZZ06}, which provides a possible
solution to the Big-Bang singularity.  An extension to the
braneworld scenario is discussed in Ref. \cite{MS09}.

The purpose of this paper is to study various wormhole
spacetimes that are supported by quintom dark energy.  We
discuss the structure of such wormholes by means of an
embedding diagram, as well as the junction to an external
Schwarzschild spacetime.  Both isotropic and anisotropic
pressures are considered.  It is shown that in the isotropic
models, the phantom-energy condition $\omega_{ph}<-1$ implies
that $\omega_q<-1$, while in the anisotropic case, $\omega_q
<-1/3$.

As a final comment, since quintessence satisfies the weak and
null energy conditions, it is not feasible to model a wormhole
using the quintessence field.  So when considered as separate
fields, only phantom energy can support a wormhole structure.

\section{{\textbf{ Construction of quintom wormholes}} }

In the  present study the metric for a static spherically symmetric
wormhole spacetime  is taken as

\begin{multline}
ds^{2}=-e^{\nu(r)}dt^{2}+e^{\lambda(r)}dr^{2}\\+r^{2}(d\theta^{2}
+\text{sin}^{2}\theta d\phi^{2}), \label{Eq3}%
\end{multline}
where $\ \nu(r)$ and $\lambda(r)$ are functions of the radial
coordinate $r$.

Let us now consider a quintom model, which, as noted earlier,
contains a quintessence-like field and a phantom-like field.
So we assume that the Einstein field equations can be written as
\begin{equation}
              G_{\mu\nu}=   8 \pi G (   T_{\mu\nu}+  \tau_{\mu\nu}),
         \label{Eq3}
          \end{equation}

where $\tau_{\mu\nu}$ is the energy-momentum tensor of the
quintessence-like field and is characterized by a free parameter
$\omega_q$, which is ordinarily restricted by the condition
$\omega_q<-\frac{1}{3}$.  (Recall that this condition is
required for an accelerated expansion.) According to
Kiselev \cite{vK03}, the components of this tensor need
to satisfy the conditions of additivity and linearity.
Taking into account the different signatures used in
the line elements, the components can be stated as
follows:
\begin{equation}
              \tau_t^t=    \tau_r^r = -\rho_q
         \label{Eq3},
          \end{equation}
\begin{equation}
              \tau_\theta^\theta=    \tau_\phi^\phi = \frac{1}{2}(
              3\omega_q+1)\rho_q
         \label{Eq3}.
          \end{equation}
The  energy-momentum tensor compatible with
          spherically  symmetry is
\begin{equation}
                T_\nu^\mu=  ( \rho + p_t)u^{\mu}u_{\nu}
    - p_t g^{\mu}_{\nu}+ (p_r -p_t )\eta^{\mu}\eta_{\nu}.
         \label{Eq3}
          \end{equation}
As already noted, a phantom-like field is
characterized by the equation of state
\begin{equation}p_r = \omega_{ph} \rho, \end{equation}
where $p_r$ is the radial pressure and
$\omega_{ph}<-1$.  In the discussion below, $p_t$ is
the lateral pressure.

The Einstein field equations in the orthonormal frame are
stated next:
\begin{equation}
e^{-\lambda}\left[
\frac{\lambda^{\prime}}{r}-\frac{1}{r^{2}}\right]
+\frac{1}{r^{2}} = 8\pi G (\rho + \rho_q),
         \label{Eq3}
          \end{equation}
\begin{equation}
e^{-\lambda}\left[  \frac{1}{r^{2}}+\frac{\nu^{\prime}}{r}\right]
-\frac {1}{r^{2}} = 8\pi G (p_r - \rho_q),
         \label{Eq3}
          \end{equation}
           \[ \frac{1}{2}e^{-\lambda}\left[
\frac{1}{2}(\nu^{\prime})^{2}+\nu^{\prime
\prime}-\frac{1}{2}\lambda^{\prime}\nu^{\prime}+\frac{1}{r}({\nu^{\prime
}-\lambda^{\prime}})\right]\]
                 \[= 8\pi G \left(p_t +\frac{(3\omega_q+1)}{2}
                 \rho_q\right).\]
       \begin{equation}   \end{equation}

\section{{\textbf{ Model 1:  A constant redshift
function}}}

For our first model we assume a constant redshift function,
  \begin{equation} \nu(r) \equiv \nu_o = \text{constant}, \end{equation}
referred to as the zero-tidal-force solution in Ref. \cite{MT88}.  The
absence of tidal forces automatically satisfies a key
traversability criterion.

\subsection{Isotropic pressure}

Our first assumption in the present model is an isotropic pressure:
\[
  p=p_r=p_t.
\]
Adding  Eqs. (7) and (8) and using Eqs. (6) and (10), we get
\begin{equation}  e^{-\lambda}\left[
\frac{\lambda^{\prime}}{r}\right] =  8\pi G(\omega_{ph}+1 )\rho.
\end{equation}
Multiplying Eq. (8) by $(3\omega_q+1)/2$ and adding
to Eq. (9) leads to
\begin{equation}   (e^{-\lambda})^{\prime} +
    \frac{A_1 e^{-\lambda}}{r}  =  \frac{A_1}{r},
\end{equation}
where
\begin{equation} A_1  =  \frac{(3\omega_q+1)(\omega_{ph}+1)}
   {(\omega_{ph}+1) +3\omega_{ph}(\omega_q+1)}.
\end{equation}

The above equation yields
\begin{equation} e^{-\lambda} =  1-\frac{D}{r^{A_1}},
\end{equation}
where $D>0$ is an integration constant. We rewrite the metric in
the Morris-Thorne canonical form \cite{MT88}, $ e^{\lambda}  =
1/[1-b(r)/r] $, where the shape function is given by
\begin{equation} b(r)  =  \frac{D}{r^{A_1-1}}.
\end{equation}

Using Eqs. (7) and (8), one gets the following forms for $\rho$
and $\rho_q$:
\begin{equation} 8 \pi G \rho=
   -\frac{DA_1}{(1+\omega_{ph})r^{A_1+2}}
\end{equation}
and
\begin{equation} 8 \pi G \rho_q  =  \frac{D(-A_1+1
   +A_1/(1+\omega_{ph}))}{r^{A_1+2}}.
\end{equation}

Observe that $\rho>0$, since $1+\omega_{ph} < 0$, while
$\rho_q >0 $ implies that $A_1<(1+\omega_{ph})/\omega_{ph}$.
It follows that $A_1<1$.

The assumption $\nu(r) \equiv \nu_0 $ implies the absence of a
horizon.  Also, we would like the wormhole spacetime to be
asymptotically flat, that is, $b(r)/r\rightarrow 0 $
as $r \rightarrow \infty$.  To this end, we require that
$A_1>0$.  From Eq. (13) we deduce that
\[
   \omega_q<\frac{-4\omega_{ph} -1}{3\omega_{ph}}.
\]
Since $\omega_{ph}<-1$, it now follows that $\omega_q<-1$, thereby
having crossed the phantom divide.  This result is hardly
surprising, given the nature of phantom wormholes.  On
the other hand, the quintessence condition $\omega_q<-1/3$
is still going to occur, namely in the anisotropic case,
discussed next.

\subsection{Anisotropic pressure}
In the case of an anisotropic pressure, the radial and lateral
pressures are no longer equal.  In an earlier paper on
phantom-energy wormholes, Zaslavskii \cite{oZ05} proposed the
form $p_t=\alpha\rho$, $\alpha>0$, for the lateral pressure.
In this manner we obtain simple linear relationships
between pressure and energy-density, but with $p_r$ not
equal to $p_t$.

Observe first that Eq. (11) remains the same.  After
multiplying Eq. (8) by $(3\omega_q+1)/2$ and adding to
Eq. (9), we get, analogously,
\begin{equation}   (e^{-\lambda})^{\prime} +
  \frac{A_2 e^{-\lambda}}{r}  =  \frac{A_2}{r},
\end{equation}
where
\begin{equation}\label{E:A1}
  A_2=\frac{(3\omega_q+1)(\omega_{ph}+1)}
  {(\omega_{ph}+1)+2\alpha+\omega_{ph}(3\omega_q+1)}.
\end{equation}
Otherwise Eqs. (14)-(17) retain their form.

As before, we want $A_2>0$.  So from Eq. (19),
we have
\[
  \omega_{ph}+1+2\alpha+\omega_{ph}(3\omega_q+1)>0
\]
and
\begin{equation}\label{E:quintessence}
  \omega_q<\frac{-2\omega_{ph}-1-2\alpha}{3\omega_{ph}}.
\end{equation}
Since $\alpha>0$ and $\omega_{ph}<-1$, it follows that
\[
     \omega_q<-\frac{1}{3},
\]
which is the condition for quintessence.  In other words, in the
anisotropic model, $\omega_q$ does not have to cross the
phantom divide.

\emph{Remark:} The parameter $\alpha$ could be negative.  For
example, if $\alpha<-1$, we return to $\omega_q<-1$.

\section{{\textbf{ Wormhole structure}} }
In this section we let $A=A_i, i=1, 2$.  Returning to the shape
function $b(r)=D/r^{A-1}$, to meet the condition $b(r_0)=r_0$,
we must have $D=r^A_0$.  So the radius of the throat is $r_0=D^{1/A}$
and
\[
   b(r)=r\left(\frac{r_0}{r}\right)^A.
\]
Since $A>0$, it now follows that $b^{\prime}(r_0)< 1$, thereby
satisfying the flare-out condition.  From Ref. \cite{MT88},
we therefore obtain the ``exoticity condition"
\begin{equation*}
  \frac{b(r_0)-r_0b'(r_0)}{2[b(r_0)]^2}=
     \frac{A}{2r_0^2}>0,
\end{equation*}
which shows that the weak energy condition has been
violated.  We already checked the asymptotic flatness,
so that our solution describes a static traversable
wormhole supported by quintom dark energy.

As discussed in Ref. \cite{MT88}, one can picture the
spacial shape of a wormhole by rotating the profile curve
$z=z(r)$ about the $z-$axis.  This curve is defined by
\begin{equation}
 \frac{dz}{dr}=\pm
 \frac{1}{\sqrt{\displaystyle{r/b(r)}-1}}= \pm
 \frac{1}{\sqrt{{r^A/D}-1}}.
 \end{equation}
For example, choosing $A=\frac{1}{2}$, we find that
\begin{equation}    z= 4\sqrt{D}   \left[ \frac{1}{3}
   ( \sqrt{r}-D)^{3/2}  +
    D\sqrt{(\sqrt{r}-D)}\right].  \end{equation}
The profile curve is shown in Figure 1 and the embedding diagram
in Figure 2.  The proper distance $l(r)$ from the throat to
a point outside is given by
\begin{figure}[htbp]
    \centering
        \includegraphics[scale=.35]{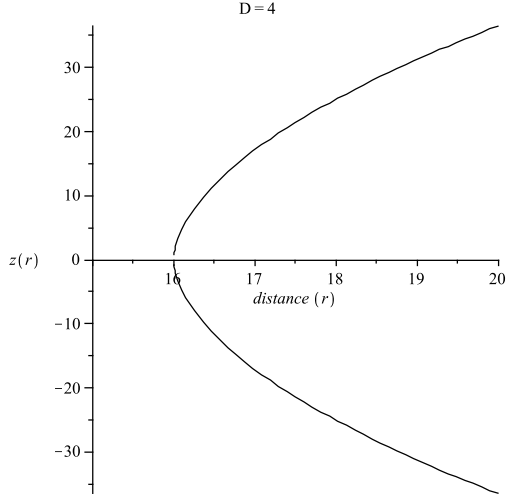}
        \caption{ The profile curve of the wormhole.}
\end{figure}

\begin{figure}[htbp]
    \centering
        \includegraphics[scale=.45]{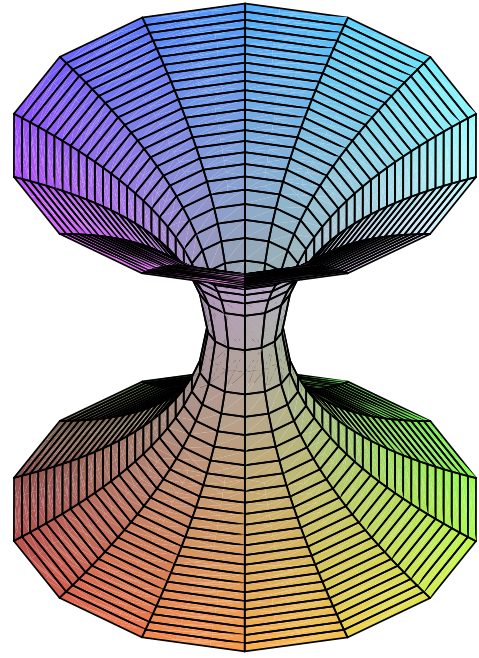}
        \caption{The embedding diagram generated by rotating
    the profile curve (fig 1) about the $z-$axis.}
\end{figure}

\begin{equation}
 l(r) = \pm \int_{r_0^+}^r \frac{dr}{\sqrt{1-b(r)/r}}.
            \label{Eq20}
          \end{equation}
 For $A=\frac{1}{2}$,
\begin{multline}
  l(r)=r^{1/4}(\sqrt{r}-D)^{3/2}+\frac{5D}{2}
   r^{1/4}(\sqrt{r}-D)^{1/2}\\
  +\frac{3}{2}D^2\text{ln}\left|
   \frac{r^{1/4}+(\sqrt{r}-D)^{1/2}}
       {\sqrt{D}}\right|.
\end{multline}
(See Figure 3.)

\begin{figure}[htbp]
    \centering
        \includegraphics[scale=.35]{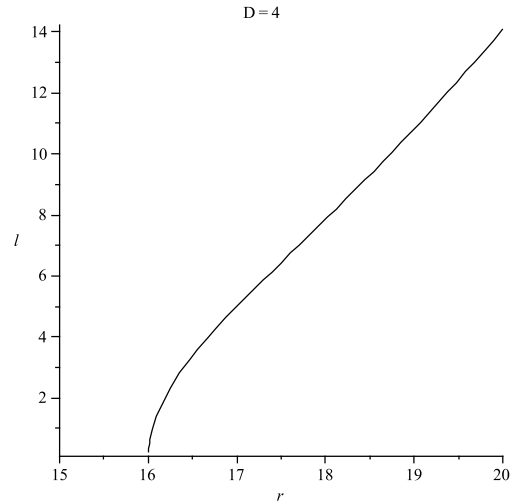}
        \caption{ The graph of the radial proper distance $l(r)$.}
\end{figure}

It is customary to join the interior solution of a wormhole to
an exterior Schwarzschild solution at some $r=a$.  To do so,
we demand that $g_{tt}$ and $g_{rr}$ be continuous at $r=a$:
\[
 {g_{tt}}_{(int)}(a) =  {g_{tt}}_{(ext)}(a)
\]
and
\[
 {g_{rr}}_{(int)}(a) =  {g_{rr}}_{(ext)}(a);
\]
$ g_{\theta\theta} $ and $ g_{\phi\phi}$ are already
continuous \cite{LLO03}.  So $ e^{\nu_0}
= 1 - \frac{2GM}{a}$ and $  1 - \frac{b(a)}{a}
 = 1 - \frac{2GM}{a}$.  This, in turn, implies that
$ D/ a^{A-1} = 2GM $.  Hence,  the matching occurs at
 \begin{equation} a =
\left(\frac{D}{2GM}\right)^{1/(A-1)}.   \end{equation}
The interior metric $ (r_0 < r \le a) $ is given by

\begin{multline}\label{Eq3}
               ds^2=  - \left[ 1- \frac{D}{ a^{A}}\right]dt^2+ \frac{dr^2}{
               1-D/ r^{A}}\\
   +r^2(d\theta^{2}
   +\text{sin}^{2}\theta d\phi^{2})
          \end{multline}

and the exterior metric $ (a < r < \infty)   $ by

\begin{multline}\label{Eq3}
               ds^2=  - \left[ 1- \frac{D}{ a^{A-1}r}\right]dt^2+
    \frac{dr^2}{1-D/ a^{A-1}r}\\
    +r^2(d\theta^{2}
   +\text{sin}^{2}\theta d\phi^{2}).
          \end{multline}






\section{ {\textbf{ Model 2:  A specific shape function}}}

Returning to the isotropic case $p=p_r=p_t$, let us eliminate
$\rho$ and $\rho_q$ in Eqs. (7) - (9) to obtain the following
master equation:

\begin{multline} \frac{1}{4}(\nu^{\prime})^{2}+
  \frac{1}{2}\nu^{\prime\prime}+\nu^{\prime} g(r)\\
  =-\frac{e^{\lambda}}{2r}\frac{(\omega_{ph}+1)
   +3\omega_{ph}(\omega_q+1)}  {\omega_{ph}+1}f(r),
\end{multline}
\\
\\
\\
where
\begin{multline} g(r) = -\frac{\lambda^\prime}{4} +\frac{1}{2r}
   +\frac{3\omega_q+1}{2r}\\
   -\frac{3\omega_{ph}(\omega_q+1)}{2(\omega_{ph}+1)r}
\end{multline}
and
\begin{equation} f(r) = (e^{-\lambda})^{\prime} +
  \frac{A_1 e^{-\lambda}}{r}  - \frac{A_1}{r}.
\end{equation}
As in Eq. (13),
\begin{equation}
 A_1  =  \frac{(3\omega_q+1)(\omega_{ph}+1)}{(\omega_{ph}+1)
    +3\omega_{ph}(\omega_q+1)}.
\end{equation}

Now we choose the shape function $b(r)$ in such a way that the
right-hand side of Eq. (28) is zero. For this specific
choice, one gets the following solution:
\begin{equation} e^{-\lambda} =  \frac{1}{1-b(r)/r} =
  1-\frac{D}{r^{A_1}},
\end{equation}
where $D>0$ is an integration constant. Fortunately, this form is
the same as in Model 1. So the physical characteristics, such as
the profile curve, the embedding diagram, and the proper radial
distance,  remain the same. But the redshift function and
stress-energy components are different.

By making the proper substitutions, one gets from Eq. (28)
\begin{equation} \nu^{\prime\prime} +\frac{ \nu^{\prime}}{r}
\left[\frac{A_1D}{2(r^{A_1}-D)}+L_1\right]
  =-\frac{(\nu^{\prime})^{2}}{2},
\end{equation}
where
\begin{equation} L_1  =
   (3\omega_q+2)-\frac{3\omega_{ph}(\omega_q+1)}
   {(\omega_{ph}+1)}.
\end{equation}

Solving this equation, we get
\begin{equation} \nu =\ln \left[EDA_1 +
   \sqrt{1-\frac{D}{r^{A_1}}}\right]^2,
\end{equation}
where $E$ is an integration constant.  We have used the
condition  $ L_1 = A_1+1$.  Starting with $L_1-1=A_1$
and simplifying, one can readily deduce that
$\omega_q=\omega_{ph}$; so once again, $\omega_q<-1$,
thereby having crossed the phantom divide.  This is
consistent with the isotropic case discussed in
Subsection 3.1.  (There is a similar consistency with
the anisotropic case, as shown at the end of the
present section.)

Finally, we get the following forms for $\rho$ and
$\rho_q$:

\begin{multline} 8 \pi G \rho  =
-\frac{DA_1}{(1+\omega_{ph})r^{A_1+2}}\times \\ \left[
\frac{A_1DEr^{A_1/2}}{\sqrt{(r^{A_1-D})}+
A_1DEr^{A_1/2}}\right]
\end{multline}
and
 \[8 \pi G \rho_q  =  \frac{D(-A_1+1)}{r^{A_1+2}}+\]
    \[\frac{DA_1}{(1+\omega_{ph})r^{A_1+2}}\left[
\frac{A_1DEr^{A_1/2}}{\sqrt{(r^{A_1-D})}+
A_1DEr^{A_1/2}}\right].\]
 \begin{equation} \end{equation}

In the anisotropic case $p_t=\alpha\rho$, $\alpha>0$,
eliminating $\rho$ and $\rho_q$ in Eqs. (7)-(9) yields
\begin{multline}
  \frac{1}{4}(\nu')^2+\frac{1}{2}\nu''+\nu'g(r)=\\
   -\frac{e^{-\lambda}}{2r}
   \frac{(\omega_{ph}+1)+2\alpha+\omega_{ph}(3\omega_q+1)}
        {\omega_{ph}+1}f(r),
\end{multline}
where
\begin{multline}
  g(r)=-\frac{1}{4}\lambda'+\frac{1}{2r}+
        \frac{3\omega_q+1}{2r}\\
    -\frac{\omega_{ph}(3\omega_q+1)
    +2\alpha}{2(\omega_{ph}+1)r}
\end{multline}
and
\begin{equation}
  f(r)=(e^{-\lambda})'+\frac{A_2e^{-\lambda}}{r}
       -\frac{A_2}{r};
\end{equation}
here $A_2$ is defined in Eq. (19).  In Eq. (33),
$A_1$ is replaced by $A_2$ and $L_1$ by
\begin{multline}
  L_2=1+(3\omega_q+1)\\
      -\frac{\omega_{ph}(3\omega_q+1)
     +2\alpha}{\omega_{ph}+1}.
\end{multline}
From the condition $L_2-1=A_2$, we deduce that
\[
   \omega_q=\frac{\omega_{ph}+2\alpha}{3}.
\]
Since $\omega_{ph}<-1$ and $\alpha>0$, we obtain, once again,
\[
     \omega_q<-\frac{1}{3}.
\]

\section{Discussion}
The combination of quintessence and phantom energy in a
joint model is referred to as quintom dark energy.  The
quintessence-like field is characterized by a free parameter
$\omega_q$ with the restriction $\omega_q<-1/3$.  For the
corresponding free parameter $\omega_{ph}$ in the
phantom-like field, the condition is $\omega_{ph}<-1$.

We have proposed in this paper that traversable wormholes
may be supported by quintom dark energy.  Two models were
considered.  The first model, a
constant redshift function, leads to the determination
of the shape function $b=b(r)$, which meets the flare-out
conditions.  The resulting spacetime is asymptotically flat.
This was followed by a brief discussion of the wormhole
structure, including an embedding diagram, proper distance,
and a junction to an external Schwarzschild spacetime.
For the second, more general model, it is possible to use
the same shape function but with a different redshift
function and stress-energy components.

In each of these models, both isotropic and anisotropic
pressures were considered.  In the isotropic case, the
phantom-energy condition $\omega_{ph}<-1$ implies that
$\omega_q<-1$, and in the anisotropic case,
$\omega_{ph}<-1$ implies that $\omega_q<-1/3$.

\subsection*{Acknowledgments} FR wishes to thank UGC,
Government of India, for providing financial support.
FR is also grateful for the research facilities
provided by IMSc.

\end{document}